\documentstyle[12pt,epsf]{article}

\catcode`@=11
\newif\if@defeqnsw \@defeqnswtrue

\def\eqnarray{\stepcounter{equation}\let\@currentlabel=\theequation
\if@defeqnsw\global\@eqnswtrue\else\global\@eqnswfalse\fi
\global\@eqnswtrue
\tabskip\@centering\let\\=\@eqncr
$$\halign to \displaywidth\bgroup\hfil\global\@eqcnt\z@
  $\displaystyle\tabskip\z@{##}$&\global\@eqcnt\@ne
  \hfil$\displaystyle{{}##{}}$\hfil
  &\global\@eqcnt\tw@ $\displaystyle{##}$\hfil
  \tabskip\@centering&\llap{##}\tabskip\z@\cr}

\def\yesnumber{\global\@eqnswtrue}

\def\@@eqncr{\let\@tempa\relax\global\advance\@eqcnt by \@ne
    \ifcase\@eqcnt \def\@tempa{& & & &}\or \def\@tempa{& & &}\or
     \def\@tempa{& &}\or \def\@tempa{&}\else\fi
     \@tempa \if@eqnsw\@eqnnum\stepcounter{equation}\fi
     \if@defeqnsw\global\@eqnswtrue\else\global\@eqnswfalse\fi
     \global\@eqcnt\z@\cr}

\def\@eqnacr{{\ifnum0=`}\fi\@ifstar{\@yeqnacr}{\@yeqnacr}}

\def\@yeqnacr{\@ifnextchar [{\@xeqnacr}{\@xeqnacr[\z@]}}

\def\@xeqnacr[#1]{\ifnum0=`{\fi}\cr \noalign{\vskip\jot\vskip #1\relax}}

\def\eqalign{\null\,\vcenter\bgroup\openup1\jot \m@th \let\\=\@eqnacr
\ialign\bgroup\strut
\hfil$\displaystyle{##}$&$\displaystyle{{}##}$\hfil\crcr}
\def\endeqalign{\crcr\egroup\egroup\,}

\def\cases{\left\{\,\vcenter\bgroup\normalbaselines\m@th \let\\=\@eqnacr
    \ialign\bgroup$##\hfil$&\quad##\hfil\crcr}
\def\endcases{\crcr\egroup\egroup\right.}

\def\eqalignno{\stepcounter{equation}\let\@currentlabel=\theequation
\if@defeqnsw\global\@eqnswtrue\else\global\@eqnswfalse\fi
\let\\=\@eqncr
$$\displ@y \tabskip\@centering \halign to \displaywidth\bgroup
  \global\@eqcnt\@ne\hfil
  $\@lign\displaystyle{##}$\tabskip\z@skip&\global\@eqcnt\tw@
  $\@lign\displaystyle{{}##}$\hfil\tabskip\@centering&
  \llap{\@lign##}\tabskip\z@skip\crcr}

\def\endeqalignno{\@@eqncr\egroup
      \global\advance\c@equation\m@ne$$\global\@ignoretrue}

\@namedef{eqalignno*}{\@defeqnswfalse\eqalignno}
\@namedef{endeqalignno*}{\endeqalignno}

\def\eqaligntwo{\stepcounter{equation}\let\@currentlabel=\theequation
\if@defeqnsw\global\@eqnswtrue\else\global\@eqnswfalse\fi
\let\\=\@eqncr
$$\displ@y \tabskip\@centering \halign to \displaywidth\bgroup
  \global\@eqcnt\m@ne\hfil
  $\@lign\displaystyle{##}$\tabskip\z@skip&\global\@eqcnt\z@
  $\@lign\displaystyle{{}##}$\hfil\qquad&\global\@eqcnt\@ne
  \hfil$\@lign\displaystyle{##}$&\global\@eqcnt\tw@
  $\@lign\displaystyle{{}##}$\hfil\tabskip\@centering&
  \llap{\@lign##}\tabskip\z@skip\crcr}

\def\endeqaligntwo{\@@eqncr\egroup
      \global\advance\c@equation\m@ne$$\global\@ignoretrue}

\@namedef{eqaligntwo*}{\@defeqnswfalse\eqaligntwo}
\@namedef{endeqaligntwo*}{\endeqaligntwo}

\newtoks\@stequation

\def\subequations{\refstepcounter{equation}%
  \edef\@savedequation{\the\c@equation}%
  \@stequation=\expandafter{\theequation}
  \edef\@savedtheequation{\the\@stequation}
  \edef\oldtheequation{\theequation}%
  \setcounter{equation}{0}%
  \def\theequation{\oldtheequation\alph{equation}}}

\def\endsubequations{%
  \setcounter{equation}{\@savedequation}%
  \@stequation=\expandafter{\@savedtheequation}%
  \edef\theequation{\the\@stequation}%
  \global\@ignoretrue}

\def\big#1{{\hbox{$\left#1\vcenter to1.428\ht\strutbox{}\right.\n@space$}}}
\def\Big#1{{\hbox{$\left#1\vcenter to2.142\ht\strutbox{}\right.\n@space$}}}
\def\bigg#1{{\hbox{$\left#1\vcenter to2.857\ht\strutbox{}\right.\n@space$}}}
\def\Bigg#1{{\hbox{$\left#1\vcenter to3.571\ht\strutbox{}\right.\n@space$}}}

\catcode`@=12

\begin{document}

\newcommand{\ev}{\mbox{events/(kg$\cdot$day)}}
\newcommand{\bsg}{\mbox{$b \rightarrow s \gamma$}}
\newcommand{\br}{\mbox{$Br( b \rightarrow s \gamma) $}}
\newcommand{\be}{\begin{equation}}
\newcommand{\ee}{\end{equation}}
\newcommand{\een}{\end{subequations}}
\newcommand{\ben}{\begin{subequations}}
\newcommand{\beq}{\begin{eqalignno}}
\newcommand{\eeq}{\end{eqalignno}}

\noindent

\font\fortssbx=cmssbx10 scaled \magstep2
\hbox to \hsize{
\includegraphics{uwlogo.ps}
\hskip.5in \raise.1in\hbox{\fortssbx University of Wisconsin - Madison}
\hfill$\vcenter{\hbox{\bf MADPH-95-874}
                \hbox{\bf hep-ph/9503283}
            \hbox{March 1995}}$ }

\vspace{1in}

\thispagestyle{empty}
\begin{center}
{\Large \bf Radiative $b$ Decays and the Detection of Supersymmetric Dark
Matter }\footnote{To appear in the Proceedings of {\it Beyond the Standard
Model IV}, Lake Tahoe, California, December 1994} \\
\vspace{7mm}
Manuel Drees\footnote{Heisenberg Fellow}\\
{\it Physics Department, University of Wisconsin, Madison, WI 53706, USA}\\
\vspace{5mm}
\end{center}

\vspace{1in}

\begin{abstract}
The upper bond on the branching ratio for \bsg\ decays implies a stringent
lower bound on the mass of the pseudoscalar Higgs boson of the MSSM if
sparticles are heavy. This leads to an upper bound on the expected event rate
in experiments searching for heavy supersymmetric dark matter. Scenarios with
lighter sparticle spectrum and light pseudoscalar Higgs boson are still
possible, but only if $\mu < 0$, which again implies a small LSP counting
rate.
\end{abstract}
\vspace*{1cm}
\clearpage
\noindent
\setcounter{footnote}{0}
\pagestyle{plain}
\setcounter{page}{1}
Recently the CLEO collaboration announced \cite{1} the observation of
inclusive \bsg\ decays:
\be \label{e1}
\br = (2.32 \pm 0.59) \cdot 10^{-4},
\ee
where the error is purely statistical. Including systematic uncertainties,
this corresponds to the bound
\be \label{e2}
\br \leq 4.2 \cdot 10^{-4}.
\ee
Since the systematic uncertainty is large, it is difficult to assign a
definite confidence level to this bound; usually it is treated as an
``effective" 95\% c.l. upper bound.

In the Standard Model (SM), this decay only proceeds via 1--loop diagrams
\cite{2}, most of which involve heavy particles ($W$ and top). The prediction
can therefore change significantly \cite{3} in extensions of the SM that
introduce new heavy particles and/or new interactions. In particular, in the
minimal supersymmetric standard model (MSSM), three new classes of diagrams
contribute \cite{4} to this decay: Diagrams involving a charged Higgs boson
and an up--type quark, diagrams with a chargino and an up--type squark, and
diagrams with a neutralino or gluino and a down--type squark; in all cases the
dominant contribution comes from third generation (s)quarks. The Higgs
diagrams always add constructively to the SM contribution, while the
contributions from sparticle loops can have either sign.

Diagrams with a gluino or neutralino in the loop are sensitive to the
difference in flavor--mixing in the quark and squark sectors, since only a
misalignment of these two sectors can generate flavor off--diagonal couplings
of the type $\tilde{g} \tilde{q}_i q_j \ (i \neq j)$. If one assumes that
squarks are degenerate at some scale, e.g. the Planck or GUT scale, such a
misalignment is itself only produced radiatively. In such models one therefore
finds \cite{4} gluino contributions to be sub--dominant, while neutralino
contributions are negligible. For simplicity I will assume here that the soft
SUSY breaking squark masses are degenerate at the weak scale; in this case the
gluino and neutralino contributions to \bsg\ decays vanish.

Both the contributions involving a charged Higgs boson and those from squarks
and charginos become small if the particles in the loop are heavy; the Higgs
contribution decouples less quickly, due to a factor $\log m^2_{H^+}/m^2_t$.
In the limit of heavy sparticles only the $W$ and Higgs contributions survive;
in this situation the bound (\ref{e2}) implies \cite{1} a lower bound on the
charged Higgs mass of the order of 300 GeV. In other words, the experimental
bound (\ref{e2}) excludes the possibility to combine a heavy sparticle
spectrum with a light Higgs sector.

This has immediate bearing \cite{8} on expectations for event rates in dark
matter detection experiments. These experiments search for relics
from the Big Bang era that might make up the dark matter whose existence has
been inferred from the velocity with which objects like gas clouds circulate
around the galaxy (so--called rotation curves). In the MSSM the lightest
supersymmetric particle (LSP) is stable. In most cases it is the lightest
neutralino $\tilde{Z}_1$, which for a wide region of parameters has a relic
density in the cosmologically interesting range \cite{6}. In the vicinity of
the solar system LSPs that have been captured by our galaxy are expected to
have a velocity $v \sim 10^{-3} c$; they can therefore only deposit a few keV
of energy in a detector, which makes their detection quite difficult.

For most detector materials the LSP--nucleus interaction is dominated by the
spin--independent (scalar) contribution to the LSP--nucleon matrix element,
since it allows the LSP to interact {\em coherently} with an entire nucleus.
In the limit of a non--relativistic LSP such interactions are \cite{7} due to
the $t-$channel exchange of the neutral scalar Higgs bosons $h^0, H^0$ of the
MSSM, as well as the exchange of squarks in the $s-$ or $u-$channel; in most
cases the Higgs exchange contributions dominate. Note that quite often the
coupling of the lighter scalar Higgs boson $h^0$ to the LSP is suppressed; the
contribution from $H^0$ exchange can therefore be important even if $H^0$ is
significantly heavier than $h^0$. The connection \cite{8} between \bsg\ decays
and LSP counting rates becomes obvious once one realizes the close connection
between the masses of the charged and heavy neutral Higgs bosons; moreover,
the same parameters that determine the masses and couplings of the charginos
appearing in SUSY contributions to \bsg\ decays also determine the mass and
couplings of the lightest neutralino, i.e. the LSP.

This relation is illustrated in Fig.~1, taken from ref.\cite{8}. Here the
solid curves show counting rates in a ${}^{76}$Ge detector \cite{9}, while the
dashed lines, which refer to the scale to the right, show the predicted \br.
The upper set of curves is for a quite heavy LSP, of about 200 GeV; the lower
set is for a much lighter sparticle spectrum, with $m_{\rm LSP} \simeq 50$
GeV.\\
\vspace{5mm}

\hfill
\epsfxsize=0.8\hsize\epsffile{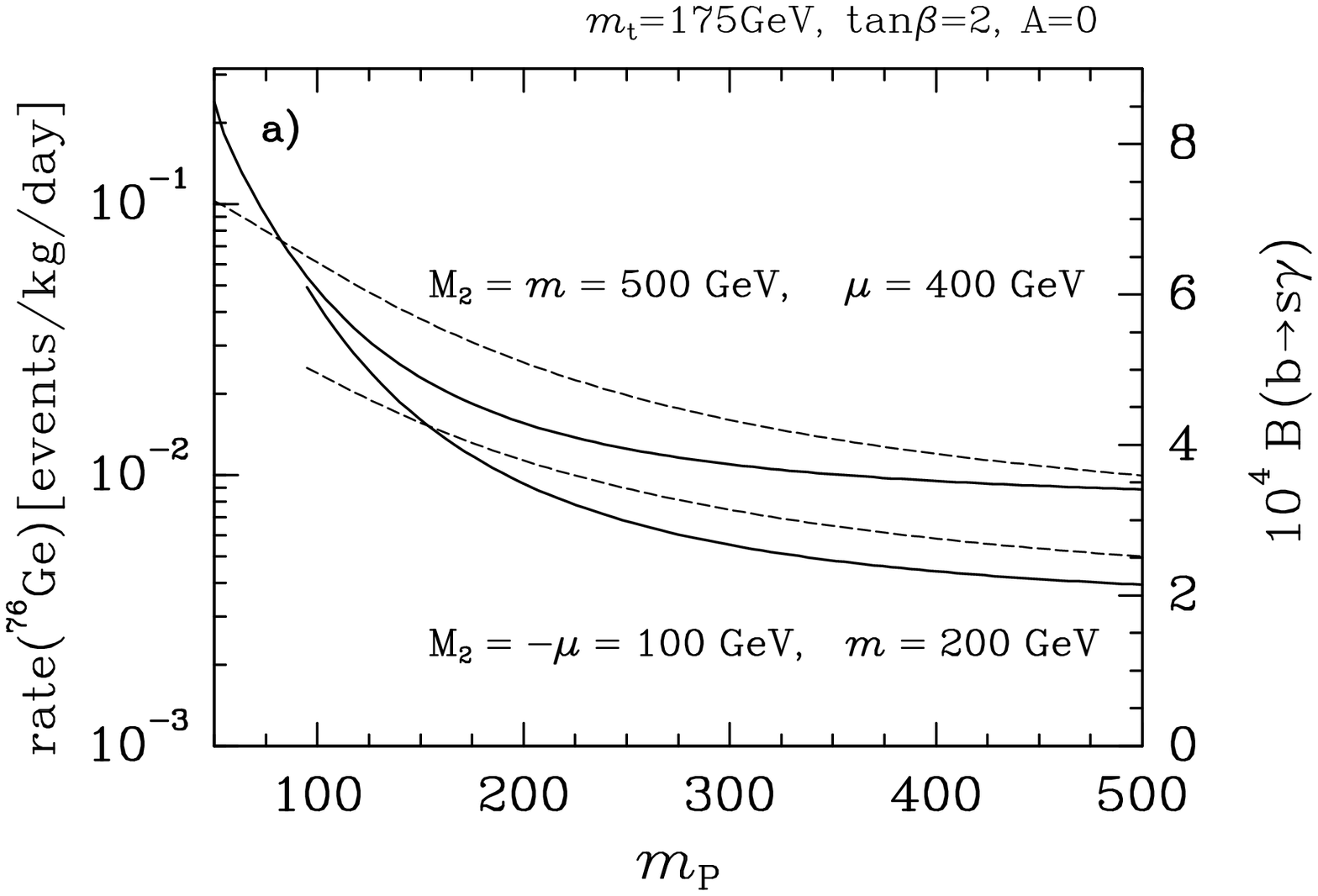} \hfill \hfill

\noindent
{\bf Fig.~1:} {\small Prediction of the dependence of the LSP counting rate in
a $^{76}$Ge detector (solid lines) and \br\ (dashed lines) on the mass $m_P$
of  the pseudoscalar Higgs boson. Results are presented  for $m_t=175$ GeV and
two different sparticle spectra, as indicated. Here, $M_2$ is the mass of the
$SU(2)$ gaugino, $\mu$ the higgsino mass parameter, and $m$ the soft breaking
squark mass.} \\
\vspace*{5mm}

We observe that both the LSP counting rate and the branching ratio for
radiative $b$ decays decrease with increasing mass $m_P$ of the pseudoscalar
Higgs boson. The branching ratio decreases since increasing $m_P$ also
increases the mass of the charged Higgs boson, and the counting rate decreases
since increasing $m_P$ raises the mass of the heavy neutral Higgs scalar.
Notice that the lower dashed curve falls below the SM expectation of $2.85
\cdot 10^{-4}$ for the branching ratio if $m_P > 350$ GeV; the reason is that
for the given choice of parameters (negative $\mu$, vanishing soft breaking
parameter $A$) the chargino--squark diagrams have the opposite sign as the
SM and charged Higgs contributions. For the heavy LSP case the contribution
from SUSY loops is almost negligible; the fact that the upper dashed curve is
well above the SM expectation even for $m_P=500$ GeV illustrates the rather
slow decoupling of the charged Higgs contributions. The bound (\ref{e2})
implies $m_P \geq 320 \ (150)$ GeV for the heavy (light) LSP scenario of
Fig.~1, which leads to an upper bound on the counting rate of 0.012 (0.015)
\ev.

Fig.~2 shows contours of constant counting rate and constant \br\ in the
$(M_2,\mu)$ plane, for both signs of $\mu$. In both cases the region
underneath the solid line is excluded by SUSY searches at colliders (region of
small $M_2$ or small $|\mu|$), or by the requirement that the lightest (stop)
squark be heavier than the lightest neutralino (small $M_2$, large $|\mu|$);
note that the soft breaking squark mass $m=2 m_{\rm LSP}$ here. Moreover, in
the areas enclosed by the dotted lines (sizable $M_2$ and $|\mu| \leq M_2/2$)
the LSP relic density is too small to be of cosmological interest; here the
LSP cannot form the dark matter in galactic haloes. Notice that requiring a
sufficiently large LSP density already excludes a substantial part of the
plane where a counting rate exceeding 0.1 \ev\ could be expected if the local
LSP density were fixed.\footnote{In fact, the region of too small relic
density is slightly bigger than shown in Fig.~2; narrow strips of low relic
density, where some $s-$channel LSP annihilation diagram becomes resonant,
have been omitted in these figures in the interest of greater clarity.}

Imposing the bound (2) further limits the available region of the $(M_2,\mu)$
plane. For $\mu > 0$, only the small area around $M_2 = 100$ GeV, $\mu = 450$
GeV below the long dashed curve survives; in particular, the entire region
where the counting rate exceeds 0.1 \ev\ is now excluded. For negative $\mu$
the much larger region below and to the right of the long dashed curve is
still viable; however, as we already saw in Fig.~1, the counting rate is
always quite small if $\mu < 0$.

Unfortunately the prediction for \br\ is still quite uncertain, mostly due to
unknown higher--order QCD corrections; experimental errors on CKM matrix
elements and the like also play a role. In ref.\cite{2} this uncertainty has
been estimated to be about $\pm 25\%$ in the SM. The same sources of
uncertainty also exist in the MSSM. The long dashed curve in Fig.~2 labelled
``low" ($\mu>0$ only) has been computed by subtracting one theoretical
``standard deviation" from the best estimate, using the formalism of
ref.\cite{2}. Only the region below this curve (at $M_2 \simeq 120$ GeV, $\mu
\simeq 200$ GeV) is in conflict with the bound (\ref{e2}) if this lower
theoretical estimate is used. However, when combined with the requirement of a
sufficiently large LSP relic density this still excludes almost the entire
region where the counting rate exceeds 1 event/(kg$\cdot$day). If this lower
theoretical estimate for \br\ is used the entire half plane with $\mu < 0$
remains viable. It should be mentioned, however, that the bound (\ref{e2}) is
based on a rather conservative treatment of the systematic uncertainties; in
particular, statistical and systematic errors have been added linearly. In my
opinion combinations of parameters which give a central estimate for \br\ that
violates the bound (\ref{e2}) while the lower theoretical estimate does not,
are therefore already strongly disfavoured, although it might be somewhat
premature to exclude them altogether.

\vspace*{5mm}
\makebox[15.2cm]{
\epsfxsize=0.75\hsize\epsffile{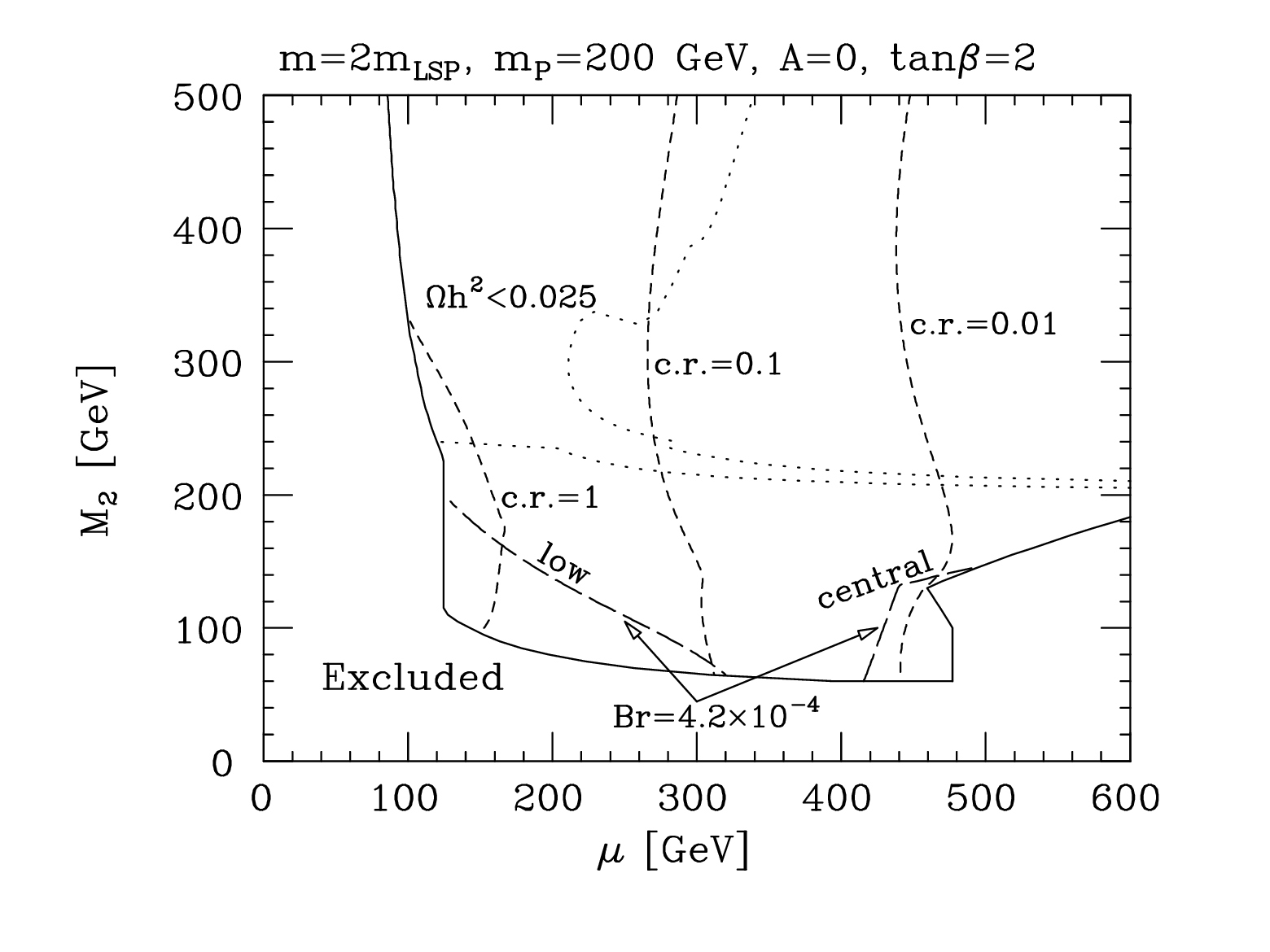} 
}

\makebox[15.2cm]{
\epsfxsize=0.75\hsize\epsffile{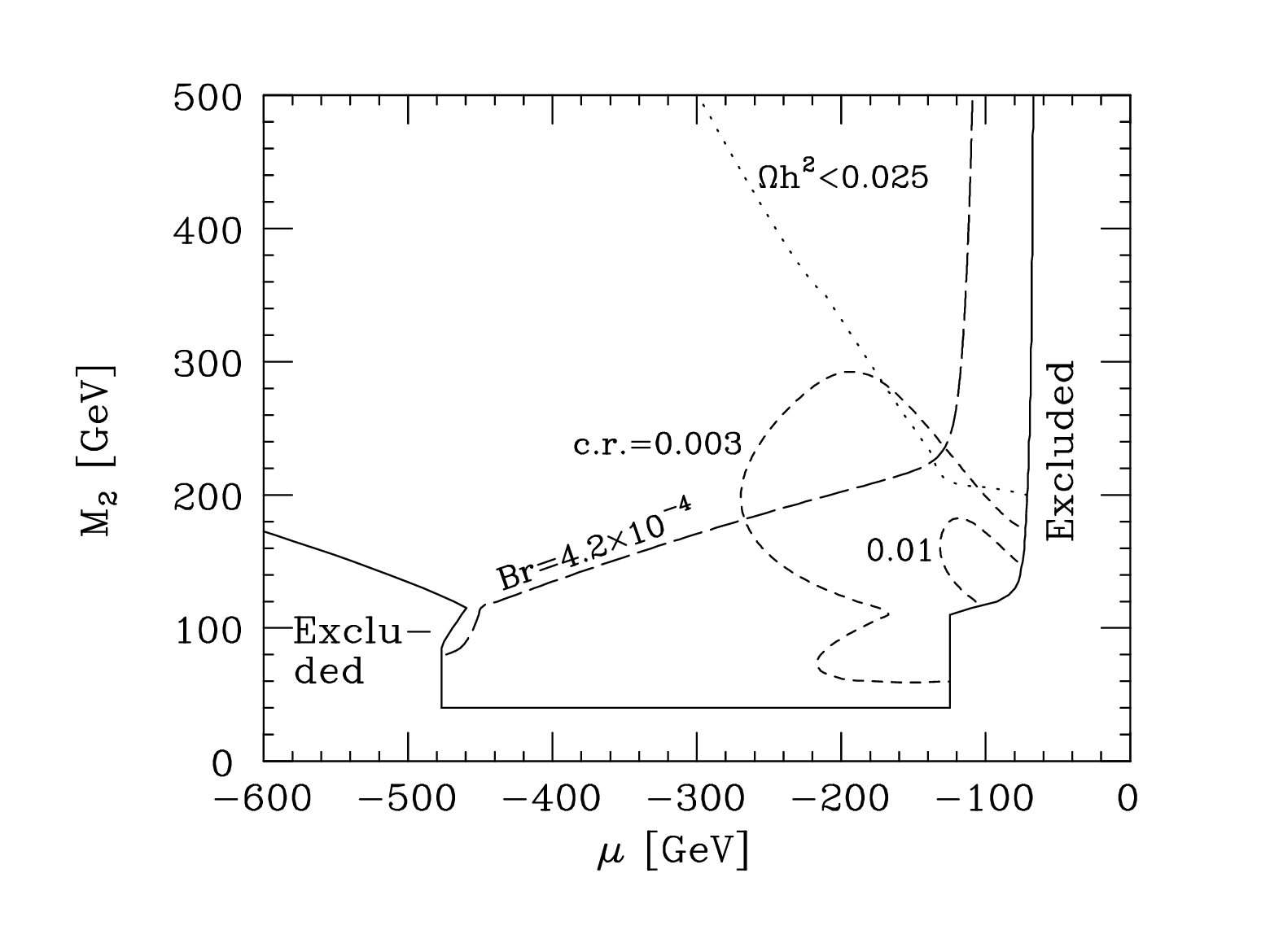} 
}

\noindent
{\bf Fig.~2:} {\small Contours of constant LSP counting rate in a $^{76}$Ge
detector [short dashed lines, in \ev] and of constant $\br=4.2 \cdot 10^{-4}$
(long dashed lines) in the $(M_2,\mu)$ plane; the two long dashed curves for
$\mu>0$ correspond to different theoretical estimates of \br, as discussed in
the text. The regions enclosed by the dotted lines have too small a relic
density for the LSP to be a good dark matter candidate.} \\

Finally, it should be mentioned that in ``minimal supergravity" models with
radiative gauge symmetry breaking a heavy sparticle spectrum more or less
automatically implies heavy pseudoscalar and charged Higgs bosons; the impact
of the bound (\ref{e2}) on the expected LSP detection rate is therefore weaker
in such models \cite{10}. Notice that these models generally predict a rather
low counting rate anyway \cite{7,10}. However, such models entail several
assumptions about physics at scales well above the weak scale. It is therefore
important to emphasize that now a purely experimental bound forces us to
expect rather low rates for dark matter search experiments if dark matter is
indeed made from superparticles.

\subsection*{Acknowledgements}
I wish to thank my collaborators Francesca Borzumati and Mihoko Nojiri;
without them, ref.\cite{8} would not have been written and I would have missed
the opportunity to have a truly cold lunch. This work was supported in part by
the U.S. Department of Energy under Grant No.~DE-FG02-95ER40896, by the
Wisconsin Research Committee with funds granted by the Wisconsin Alumni
Research Foundation, as well as by a grant from the Deutsche
Forschungsgemeinschaft under the Heisenberg program.


\end{document}